\documentclass[journal=jacsat,manuscript=article,amsmath,amssymb,aps,floatfix,superscriptaddress]{achemso}

\usepackage{nicefrac,amsmath,amssymb,mathrsfs}
\usepackage{graphicx}
\usepackage{natbib}
\usepackage{bm,color}
\usepackage{xspace}
\usepackage{tabularx}
\usepackage{hyperref}
\usepackage{verbatim}
\usepackage{appendix}
\usepackage{enumitem}
\usepackage{mathtools}
\usepackage{siunitx}
\DeclareSIUnit{\pixel}{px}

\title{Low-energy single-electron detector with sub-micron resolution}

\author{Luis Alfredo Ixquiac Méndez}
\affiliation{University of Vienna, Faculty of Physics, VCQ, 1090 Vienna, Austria}
\alsoaffiliation{University of Vienna, Max Perutz Labs,
1030 Vienna, Austria}

\author{Martino Zanetti}
\affiliation{University of Vienna, Faculty of Physics, VCQ, 1090 Vienna, Austria}
\alsoaffiliation{University of Vienna, Max Perutz Labs,
1030 Vienna, Austria}

\author{Tilman Kraeft}
\affiliation{University of Vienna, Faculty of Physics, VCQ, 1090 Vienna, Austria}
\alsoaffiliation{University of Vienna, Max Perutz Labs,
1030 Vienna, Austria}

\author{Thomas Juffmann}
\email{thomas.juffmann@univie.ac.at}
\affiliation{University of Vienna, Faculty of Physics, VCQ, 1090 Vienna, Austria}
\alsoaffiliation{University of Vienna, Max Perutz Labs,
1030 Vienna, Austria}

\date{\today}

\begin{document}

\begin{abstract}

Single-electron detectors are a key component of electron microscopes and advanced electron optics experiments. 
We present a YAG:Ce scintillator-based single-electron detector with a spatial resolution of $\SI{0.9}{\mu m}$ at an electron energy of \SI{17}{keV}. Single-electron detection events are identified with an efficiency and purity larger than 0.8 at an electron energy of \SI{17}{keV}, reaching 0.96 at \SI{30}{keV}. We show that the detector enables electron diffraction studies with a sample-detector distance comparable to the mean free path of electrons at atmospheric pressure, potentially enabling atmospheric electron diffraction studies. 

\end{abstract}

\section{Introduction}

Electron detectors are a crucial component of electron microscopes, electron spectroscopy setups, and quantum electron optics experiments. High spatial and temporal resolution, low noise, and single electron detection efficiency are among the key features of modern electron detectors. 

The high sensitivity and speed of direct electron detectors (DED) ~\cite{mcmullan_chapter_2016} have revolutionized dose-sensitive applications such as cryo-electron microscopy or tomography~\cite{glaeser_how_2019}, and DEDs are now the gold standard in most electron microscopy and spectroscopy applications~\cite{levin_direct_2021}. At low electron energies typical for scanning electron microscopy (SEM, 1 to 30\,keV), DEDs are often employed in the form of hybrid array detectors. While they enable single electron detection, their spatial resolution is limited by their pixel size which is typically 55\,$\mu$m~\cite{llopart_timepix4_2022} or larger~\cite{philipp_very-high_2022}. Monolithic active pixel sensors offer a pixel size down to 5\,$\mu$m~\cite{wilkinson_direct_2013, wang_electron_2021, CLABBERS2022107886} and have been employed at energies down to \SI{4}{keV} \cite{wang_electron_2021}. While they are commercially available, they are often prohibitively expensive.   

Here, we demonstrate single electron detection and counting based on a YAG:Ce scintillator~\cite{CryturYAGCe} that is imaged with an optical microscope. Using a high numerical aperture objective, we collect an average of 26 photons per \SI{30}{keV} electron, yielding an efficiency and purity in classifying single-electron events of 0.96. We demonstrate single electron detection in an energy range between 17\,keV and 30\,keV, obtaining a spatial resolution of $\SI{0.9}{\mu m}$ ($\SI{2.3}{\mu m}$) at \SI{17}{keV} (\SI{30}{keV}), respectively. This is $5\times$ better than state-of-the-art direct electron detectors at the same electron energy. Finally, we show that our new detector enables electron diffraction studies at sub-mm distances between the sample and the screen. This potentially enables miniature diffraction and spectroscopy setups, as well as diffraction studies at atmospheric pressure, avoiding the need for transferring the samples into vacuum.

\section{Setup}

\begin{figure}[]
    \includegraphics[width=0.6\columnwidth]{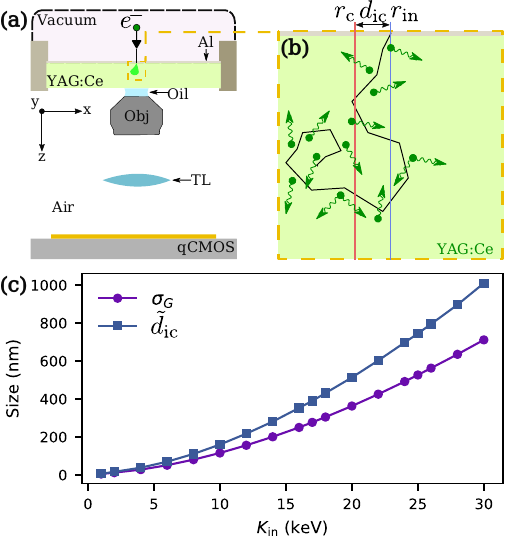} 

    \caption{ (a) Schematic of the setup: an incident electron $e^-$ hits the YAG:Ce scintillator, where a part of its kinetic energy is converted into luminescent photon emission. The resulting signal is relayed by an infinity-corrected optical system onto the CMOS camera sensor. (b) Zoomed-in illustration of the electron trajectory in the YAG:Ce scintillator (black solid line) and the emitted photons (green arrows): $r_{\mathrm{in}}$ 
    and $r_{\mathrm{c}}$ denote the coordinates of the electron incidence point and of the center of deposited energy, respectively. The distance between them is $d_{\mathrm{ic}}$. (c) Radius $\sigma_G$ enclosing $68\%$ of the emitted photons (purple line, round markers), and median distance $\tilde{d}_{\mathrm{ic}}$ between $r_{\mathrm{in}}$ and $r_{\mathrm{c}}$ (blue line, square markers), both shown as a function of the electrons' initial kinetic energy.}

    \label{fig:setup}
\end{figure}

The new detector is sketched in Figure~\ref{fig:setup} (a). Electrons from a modified FEI XL30 Scanning Electron Microscope (SEM) hit an Yttrium Aluminum Garnet scintillator doped with Cerium (YAG:Ce, Crytur \cite{CryturYAGCe}). The scintillator is 200~$\mu$m thin. 
The electrons deposit energy in the material leading to scintillation light at wavelengths around $\lambda=550 \, \si{\nano\meter}$.
To enable high-efficiency light-collection, the scintillator is imaged from behind using an oil-immersion objective of high numerical aperture (Olympus 40X UPlanXApo, NA=1.4). Similar to the design in~\cite{juffmann_real-time_2012}, the scintillator serves as a window for the vacuum chamber, allowing the objective to be mounted in air.
The oil immersion ($n_{oil}=1.51$) is crucial as it significantly increases the critical angle beyond which light is trapped within the scintillator material ($n_s=1.82$).
To further increase light collection efficiency, the scintillator is coated with a  $25\, \si{\nano\meter}$ Aluminium layer on the vacuum side, which acts as a mirror for the scintillation light (reflectance $R=0.84$ at $\lambda=550 \, \si{\nano\meter}$ \cite{FilmetricsReflectanceCalculator}). 
Overall, we expect a light collection efficiency of $\eta_L=0.38$, assuming isotropic emission from the scintillator (see the Supporting Information (SI)).

The objective and a tube lens (Thorlabs AC254-050-A-ML) form an infinity-corrected system that images the scintillation light onto a CMOS camera (Hamamatsu Orca Quest, pixel width $\SI{4.6}{\mu m}$) at a measured effective magnification of $\mathrm{M}=11.4\, \mathrm{X}$. To minimize read noise, the camera is cooled to $-34 \, ^\circ\mathrm{C}$, and operated in photon number resolving mode, yielding a read noise of  0.13 counts rms (see the SI).

\section{Simulation}

When an electron hits the scintillator, it deposits its kinetic energy through multiple collisions\cite{andreo2017fundamentals, podgorsak2016radiation}, leading to a random electron trajectory as sketched in Figure~\ref{fig:setup}b, and to the emission of scintillation light facilitated by the Ce dopants~\cite{knoll2010radiation, birks1964scintillation}. 

We perform Monte Carlo simulations using the CASINO simulation software \cite{demers2011three} to better understand the consequences of these random trajectories on our detector (see the SI). We first calculate the transverse radius $\sigma_G$ into which $68\%$ of the energy is deposited, which sets a lower bound on the point-spread function of our detector. The purple line in Figure~\ref{fig:setup}c shows  $\sigma_G$ as a function of the kinetic energy $K_{in}$ of the incoming electron. We see that it increases with $K_{in}$, but it remains below \SI{1}{\um} for energies below \SI{30}{keV}.
Next, we simulate the median distance $\Tilde{d}_{ic}$ between the transverse position $r_{in}$ at which the electron enters the scintillator and the center of deposited energy $r_{c}$ (blue line in Figure~\ref{fig:setup}c). Again, we see a non-linear increase with $K_{in}$ with a maximum value of $\Tilde{d}_{ic}=1009 \, \si{\nano\meter}$ at \SI{30}{keV}. This distance limits the accuracy for localizing single-electron detection events.

\section{Results}\label{sec:results}

\begin{figure*}[]
    \centering
    \includegraphics[width=\columnwidth]{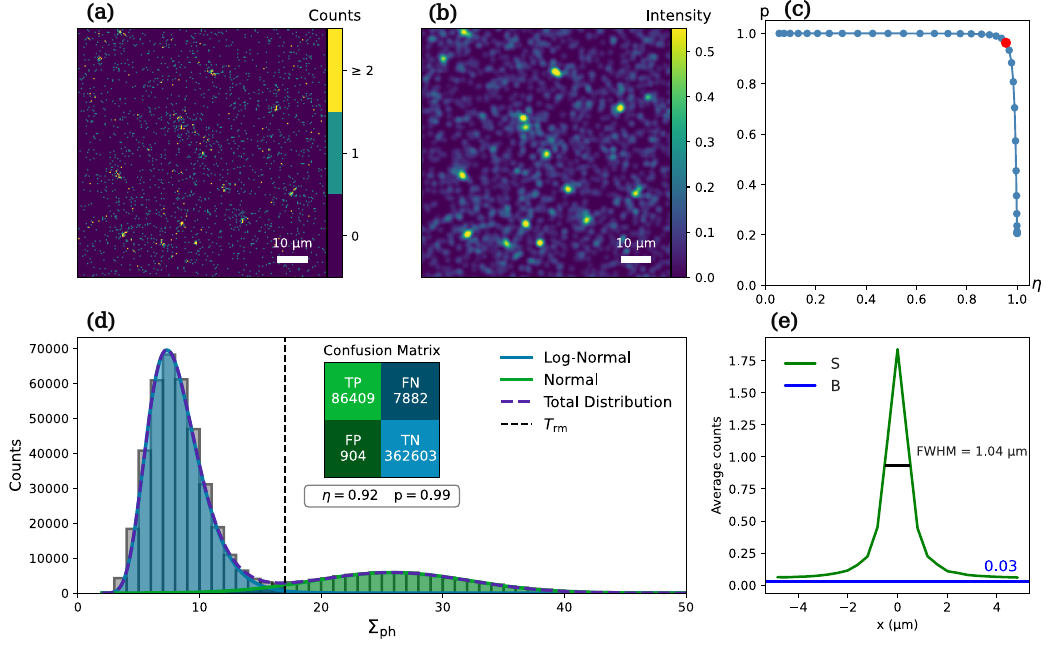} 
    \caption{Data analysis for $30\,\si{\kilo\electronvolt}$ electrons: (a) Zoom-in of a square subregion of the raw data frame; (b) Same subregion after applying a Gaussian filter. (c) purity $p$ versus efficiency $\eta$ curve, with the maximum $F_{1}$-score indicated.  (d) Histogram of photon counts within a circle of diameter $d_{\mathrm{ref}}$, centered on local maxima in the Gaussian-filtered image. The histogram is fit with a linear combination of Log-Normal and Normal distributions. The confusion matrix for binary classification is calculated with a threshold $T_{\mathrm{rm}}=17\,\mathrm{counts}$, which optimizes the purity and is used to select events for computing the average PSF. (e) Cross section of the average PSF (S, green line) with corresponding FWHM~=~1.04~$\mu$m, and the average background (B, blue line).}
    \label{fig:histogram}
\end{figure*}

\textbf{Detection of single \SI{30}{keV} electrons:} Figure \ref{fig:histogram}a shows raw data for the detection of \SI{30}{keV} electrons. While read noise leads to a random distribution of single photon counts, the electron beam induces localized detection events. To identify these events, we first subtract an averaged background image from the raw data, which we recorded with the electron beam off. 
We then apply a Gaussian filter of radius $\sigma_G$, yielding the image shown in Figure~\ref{fig:histogram}b, which shows distinct event detection candidates (details in the SI). To decide which of them correspond to single-electron detection events, we identify the local maxima in the image and calculate the total number of detected photons $\Sigma_{ph}$ within a circle of diameter $d_{ref}=\SI{5.2}{\mu m}$ (corresponding to 13 pixels) centred at each maximum. We ignore spurious events and events close to dead pixels of the camera (details in the SI). 
This yields the histogram in Figure~\ref{fig:histogram}d, which shows two distinct peaks. The one to the left, at lower values of $\Sigma_{ph}$, is due to read noise and is also present in individual background frames (see the SI). Empirically, we find that it can be fitted with a log-normal distribution. The one at higher values of $\Sigma_{ph}$ corresponds to single-electron detection events. It can be fitted with a normal distribution with a mean photon number of $\overline{\Sigma_{ph}}=26$. 

We can now use the fitted distributions to find a threshold that optimally discriminates between noise and single-electron detection events. For a given threshold, we calculate the confusion matrix, i.e., true ($T$) and false ($F$) positives ($P$) and negatives ($N$) (see the SI). The efficiency $\eta=TP/(TP+FN)$ describes the probability that an electron is correctly classified, while the purity $p=TP/(TP+FP)$ gives the ratio of detected events that actually correspond to an electron. 
Choosing a classification threshold involves a compromise between $\eta$ and $p$, as indicated in Figure~\ref{fig:histogram}c. 
It is a common choice to find the compromise by maximizing the $F_{\mathrm{1}}$-score, $F_{\mathrm{1}} \coloneq \frac{2 \eta p}{\eta+p}$. 
In our case, this maximization yields $F_{\mathrm{1}}=0.96$ for a threshold at $\Sigma_{ph}=15$, which corresponds to an efficiency $\eta=0.96$, and a purity $p=0.96$, as indicated by the red dot in Figure~\ref{fig:histogram}c.

To further characterize the optical detection scheme, we sum up 89275 detection events with their local maxima superposed, yielding a proxy for the point-spread function (PSF) of the optical system. To minimize the influence of FP events on the PSF, we choose a threshold of $\Sigma_{ph}=17$ (vertical dashed line in Figure~\ref{fig:histogram}c), corresponding to $p=0.99$. The PSF cross section is shown in Figure~\ref{fig:histogram}e (S, green line), together with the average background (B, blue line), computed as the average across all pixels of the averaged background image. 
We obtain a full width at half maximum FWHM$=\SI{1.04}{\mu m}$, assuming linear interpolation between pixel values, see Figure~\ref{fig:histogram}e. To get an estimate of the spatial resolution of our detector at \SI{30}{keV}, we also have to consider the random walk-off $\Tilde{d}_{ic}$ discussed previously. Adding them in quadrature, yields a resolution estimate of $\delta x=\sqrt{\mathrm{FWHM}^2+(2\Tilde{d}_{ic})^2}=2.3 \, \si{\micro\meter}$ 

\begin{figure}[]
    
    \includegraphics[width=0.6\columnwidth]{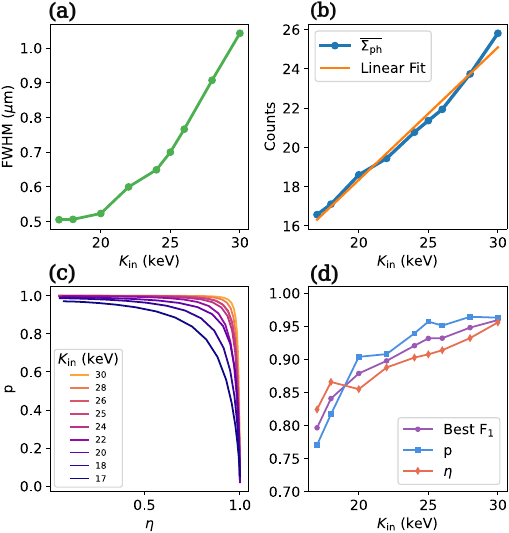} 
    \caption{(a) Measured FWHM of the PSF as a function of energy. (b) Average photon counts per detected event as a function of energy (blue line, round markers) and linear fit (orange line). The linear fit has a slope of 0.67\,photons/keV and an intercept of 4.8\,photons. (c) purity-efficiency curves for electron energies between 17\,keV and 30\,keV. (d) Best $F_{\mathrm{1}}$-score and corresponding purity $p$ and efficiency $\eta$ as a function of electron energy.}
    \label{fig:many_en}

\end{figure}

\textbf{Detection characteristics as a function of electron energy:}
Figure~\ref{fig:many_en} illustrates the detector's characteristics as a function of the electron energy $K_{in}$. First, Figure~\ref{fig:many_en}a shows that the measured FWHM increases with $K_{in}$, which is due to the increased size of the scintillation plume $\sigma_G$. At energies below \SI{20}{keV} the curve levels off, mainly due to the effective pixel size in the scintillator plane ($\SI{0.4}{\mu m}$). If we again combine this measurement with the simulated walk-off from Figure~\ref{fig:setup}c, we obtain a resolution estimate of $\delta x=0.9 \, \si{\micro\meter}$ at an electron energy of $\SI{17}{keV}$.

This increased spatial resolution comes at the cost of a higher classification error due to the lower number of photons detected per incoming electron.
Figure~\ref{fig:many_en}b shows the dependence of the mean photon number $\overline{\Sigma_{\mathrm{ph}}}$ on $K_{in}$. A linear fit yields a slope of 0.67 detected photons per keV, and an intercept of 4.8 photons, in good agreement with the measured $\overline{\Sigma_{\mathrm{ph}}}$ of the background in the SI. 

The lower number of photons per event at lower energies leads to slightly decreased sensitivity and purity, as shown in Figure~\ref{fig:many_en}c. The best $F_1$ score at each energy is shown in Figure~\ref{fig:many_en}d, along with the respective efficiency and purity. As an example, at $K_{in}=\SI{17}{keV}$, we obtain $F_1=0.8$. Note that in practice, a maximal $F_1$ score does not necessarily represent the optimal condition for a specific experiment.

\begin{figure}[]
    \includegraphics[width=0.6\columnwidth]{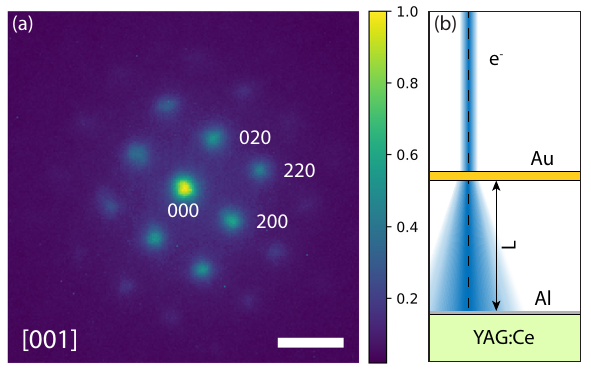} 
    \caption{(a) Diffraction pattern of a [001]-oriented single gold crystal, obtained with a 30\,keV electron beam. Scale bar: \SI{15}{\um} in the detection plane. (b) Measurement scheme: The gold crystal (Au) is on a 300 mesh gold TEM grid (omitted in sketch) and placed on top of the YAG:Ce scintillator with a plastic spacer in between, leading to a sample-screen distance of $L= \SI{380}{\mu m}$.
    }
    \label{fig:diffraction_pattern}
\end{figure}

\textbf{Towards electron diffraction at atmospheric pressures:} Lastly, we use our detector for diffraction studies in which the sample-detector distance has to be minimized. Figure~\ref{fig:diffraction_pattern} shows the diffraction pattern obtained with \SI{30}{keV} electrons from a standard oriented gold crystal (Edge Scientific, EM-Tec TC1) usually employed for transmission electron microscopy calibration. 

From the positions of the diffraction peaks, we calculate the distance between the crystal and the screen to be $L=\SI{380}{\mu m}$, comparable to the mean free path length of \SI{30}{keV} electrons in air ($\sim 75\,\mu$m) or helium ($\sim 800\,\mu$m)~\cite{nguyen_spatial_2016}.
Fitting the diffraction orders with a Gaussian yields a FWHM of $\SI{4.2}{\mu m}$. This is slightly larger than the resolution estimate $\delta x$, likely due to finite coherence, electromagnetic stray fields, and mechanical vibrations in our setup, where the detector is mounted at a distance of \SI{0.7}{m} below the objective lens of the SEM~\cite{mihaila_transverse_2022}.
Nevertheless, the distance between two diffraction orders can be determined with a precision much better than $\delta x$. Specifically, using 2D Gaussians to determine the positions of the diffraction peaks, we can measure the distance between the $0^{\mathrm{th}}$ diffraction order and the 020-peak with a precision \SI{60}{nm}, given by the standard deviation from 25 measurements. 
This corresponds to an angular precision of $\SI{160}{\mu rad}$.

\section{Discussion and Conclusion}\label{sec:summary}

We have demonstrated a single-electron detector with an energy-dependent spatial resolution on the order of $\SI{1}{\mu m}$ in the energy range 17\,keV - 30\,keV. The detector offers high efficiency and purity in distinguishing single-electron detection events from background noise. 
We have shown that the detector enables electron diffraction studies with a distance between the sample and the screen as short as $\SI{380}{\mu m}$. This is on the order of the mean free path of \SI{30}{keV} electrons in air ($\sim\SI{75}{\mu m}$), and Helium ($\sim\SI{800}{\mu m}$)~\cite{nguyen_spatial_2016}. Combined with vacuum-sealed electron guns~\cite{han_atomically_2016}, our detector thus enables diffraction studies at atmospheric pressures, complementing atmospheric scanning (transmission) electron microscopy~\cite{han_atomically_2016, nguyen_spatial_2016}. This potentially enables high-throughput studies and quality-control applications in which the samples no longer have to be transferred into vacuum.
Importantly, we demonstrated that the narrow point-spread function of the imaging system enables high precision in localizing the diffraction orders. Specifically, we showed that we can determine the angle between diffraction orders with a precision of $\SI{160}{\mu }$rad. This enables high-precision miniature diffraction and spectroscopy applications~\cite{khursheed_add-transmission_2003, shiloh_quantum-coherent_2022, tripathi_resolution_2019}. 

The use of a scintillator with custom optical detection setups enables experimental flexibility in terms of detector specifications. For example, our detector would be compatible with event-based cameras~\cite{gallego_event-based_2022}, which could enable fast acquisition at a low data acquisition rate. Considering the fast temporal response of the YAG:Ce scintillator material (rise time down to \SI{1}{ns}, decay time down to \SI{85}{ns}~\cite{zapadlik_engineering_2022}), our detector can also be an excellent choice for time-resolved studies, especially when combined with fast camera technology, such as gated intensifiers or fluorescence lifetime imaging cameras~\cite{chen_modulated_2015, bowman_electro-optic_2019, marchand_super-resolution_2025}. Future implementations might also use GAGG(Ce) scintillators, which offer higher photon yield and material density, potentially leading to a higher signal-to-noise and a smaller detection plume, respectively. 

The high resolution of our detector also lowers the demand for magnification in electron optical setups. This will enable more sensitive quantum electron optics experiments, such as ponderomotive electron wavefront shaping in a low-intensity limit~\cite{mihaila_transverse_2022}. It can also benefit compact electron optics setups, such as tabletop low-energy TEMs~\cite{drummy_electron_2014, dazon_comparison_2019}.

\begin{acknowledgement}
We thank Peter Kunnas and Philipp Haslinger for fruitful discussions. We thank the Electron
Microscopy Facility of the Vienna BioCenter Core Facilities GmbH (VBCF), member of the Vienna
BioCenter (VBC), Austria, for providing TEM calibration samples. We thank the University Service Centre for Electron Microscopy (USTEM) at the TU Wien and Isobel Bicket for TEM measurements of the gold sample and for fruitful discussions.

This project has received funding from the European Union's Horizon 2020 research and innovation program under grant agreement N.101017902.
\end{acknowledgement}

\newpage

\begin{suppinfo}

\section{Collection efficiency of the optical system}\label{sec:appendix_collect-eff}

To calculate the efficiency of the optical system, note that due to space constraints, a 45$^o$ mirror (Thorlabs PF10-03-P01P) is inserted in the optical path, and the camera is placed at 90$^o$ to the objective.
The overall efficiency of the optical system is
\begin{equation*}
    \eta_{L} = \eta_{coll}\cdot(1+R)\cdot\eta_{obj}\cdot\eta_{mirr}\cdot\eta_{T.L.}\cdot DQE
\end{equation*}
where $\eta$ are respectively the collection efficiency ($\eta_{coll}=\frac{\Omega}{4\pi} \sim$ 0.31, for NA=1.40), the objective transmission efficiency ($\eta_{obj}=0.90$), the 45$^o$ mirror reflectance ($\eta_{mirr}=0.90$) and the transmittance of the tube lens ($\eta_{T.L.}=0.97$). Furthermore, R=0.84 is the reflectance of the Aluminium layer deposited on the scintillator~\cite{FilmetricsReflectanceCalculator}, and DQE=0.8 is the detector quantum efficiency of the camera. All the parameters are estimated at the scintillation wavelength of 547\,nm, resulting in $\eta_{L}=0.38$.

\section{Monte Carlo simulation}\label{sec:appendix_monsim}

The Monte Carlo simulation is performed using the CASINO simulation software \cite{demers2011three}.
The simulation models $7 \times 10^5$ electrons with kinetic energy of $30 \, \si{\kilo\electronvolt}$ impinging at a right angle on the Al coating surface on top of the scintillator. The incident electrons move in the $+z$ direction.
For each electron, the simulation yields a trajectory of N points $\left( (x_i,y_i,z_i) : {i \in \{1, \ldots,N\}}\right)$, with corresponding kinetic energies $E_{i}$. For each electron trajectory, we calculate the center of deposited energy as
$r_{\mathrm{c}} = (x_{\mathrm{c}},y_{\mathrm{c}}) \coloneq \frac{\sum_{i=2}^{N} (x_i,y_i)\cdot(E_{i-1}-E_i) }{E_1}$

Figure~\ref{fig:monsim}a(b) shows the deposited energy, integrated along the z(y) axis. The distribution represents the average across all simulated trajectories, normalized to the total energy. Note that the distributions of individual electrons are aligned with respect to their centers of deposited energy $r_{\mathrm{c}}$. The solid lines denote the areas that contain $25$, $50$, $68/75$, and $90\%$ of the total deposited energy in a/b, respectively. These calculations are repeated for different energies, providing the data for the purple line in Figure \ref{fig:setup}c. 

Figure~\ref{fig:monsim}c shows a histogram of the distances between the coordinate of the incoming electrons and their simulated $r_{\mathrm{c}}$ for two different energies. The median of the distributions is indicated by the vertical dashed lines. These calculations are repeated for different energies, providing the data for the blue line in Figure \ref{fig:setup}c.

\begin{figure}[]
    \includegraphics[width=0.6\columnwidth]{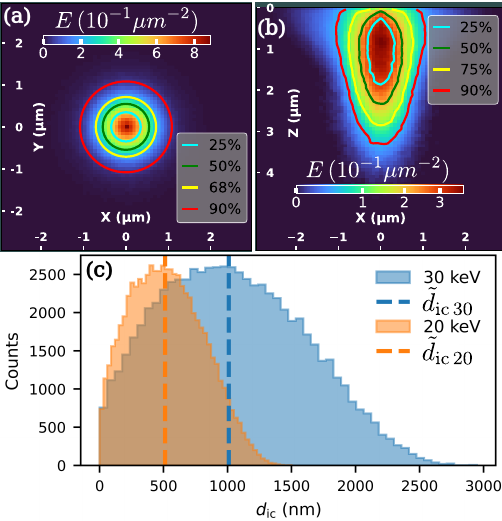} 

    \caption{Monte Carlo simulation: (a) and (b) normalized x-y and y-z projections of the spatial distribution of energy deposited by the electrons on the scintillator. Electron trajectories used in the computation are aligned with respect to their centre of deposited energy $r_{\mathrm{coel}} = (x_{\mathrm{c}},y_{\mathrm{c}})$. (c) Distributions of $d_{\mathrm{ic}}$ (distance of the center of deposited energy from the entrance point) for $30 \, \si{\kilo\electronvolt}$ electrons (blue, right) and $20 \, \si{\kilo\electronvolt}$ electrons (orange, left). The dotted lines mark the medians of each distribution.}

    \label{fig:monsim}
\end{figure}

\section{Image processing and event detection}\label{sec:appendix_imageprocess}

Image analysis is performed on a circular region of interest (ROI) with a diameter of $\SI{320}{\mu m}$.

To obtain a background image, we average 200 images with the electron gun off. 
From the same set of images, we calculate the standard deviation in background counts on each pixel.
Pixels above the $99.99 \, \mathrm{th}$ percentile of the mean and standard deviation are considered dead pixels. After background subtraction, Gaussian filtering is performed. We use the radius of the distribution of simulated deposited energy ($\sigma _{G}$, see Figure~\ref{fig:setup}c) as the width of the Gaussian, defined on a circular kernel of diameter $2 \left(\mathrm{round}(2\sigma _{G}/(\SI{0.4}{\mu m}))\right)+1 \, \si{\pixel}$ .

After finding all local maxima, we reject those that are closer than $d_{ref}/2$ to a dead pixel. Similarly, if two local maxima are at a distance smaller than $d_{ref}$, the weaker is rejected. Furthermore, we reject spurious detection events, defined as maxima with more than 7 counts. These are statistically very unlikely from single-electron events.  

\section{Characterizing Background Noise}\label{sec:appendix_bgnoise}

Performing the above procedure on the data set of background images, we obtain the histogram with the photon number distribution of dark counts shown in Figure \ref{fig:noisemod}. We found empirically that they follow a log-normal distribution.

\begin{figure}[]
    \includegraphics[width=0.6\columnwidth]{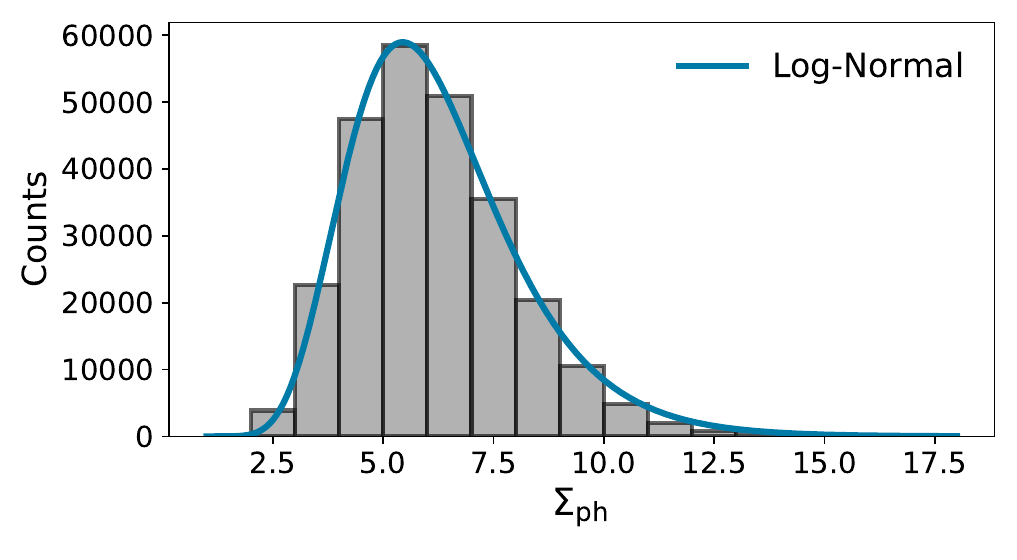} 

    \caption{Background photon number distribution: histogram of candidate events with $\Sigma_{ph}$ photons contained in a circle of diameter $d_{ref}=5.2\,\mu$m centered at local maxima coordinates of gaussian filtered images of dark background, and Log-Normal fit of the distribution.}
    
    \label{fig:noisemod}
\end{figure}

The read noise is calculated on a set of 740 images taken with the gun off across the whole chip of the camera. We calculate the standard deviation in the number of detected photons for each pixel throughout the images, and calculate the  average standard deviation across all pixels. This yields 0.13 counts rms.

\section{Binary classification}\label{sec:appendix_binclass}
With the electron gun on, we obtain the histogram in Figure~\ref{fig:histogram}d. Binary classification requires defining a threshold $T_{\mathrm{rm}}$ of photon counts to select single-electron detection events ($\Sigma_{ph}\geq T_{\mathrm{rm}}$) and reject the dark counts ($\Sigma_{ph}<T_{\mathrm{rm}}$).
 
The true positives (TP), false positives (FP), true negatives (TN), and false negatives (FN) are defined on the fitted distributions and are given by 
$$TP \coloneq \int_{T_{\mathrm{rm}}}^{\infty}N(x) dx \quad FP \coloneq \int_0^{T_{\mathrm{rm}}}N(x) dx$$
$$TN \coloneq \int_0^{T_{\mathrm{rm}}}L_{\mathrm{N}}(x) dx \quad  FN \coloneq \int_{T_{\mathrm{rm}}}^{\infty}L_{\mathrm{N}}(x) dx$$

where $N$ and $L_{\mathrm{N}}$ are, respectively, the fitted normal and log-normal distributions. These four numbers define the confusion matrix of the binary classification, which is shown in Figure \ref{fig:histogram}d.

\end{suppinfo}

\bibliography{references}

\end{document}